\documentclass[sigconf]{acmart}
\usepackage{float}
\usepackage{placeins}
\AtBeginDocument{%
  }

\setcopyright{acmlicensed}
\copyrightyear{2018}
\acmYear{2018}
\acmDOI{XXXXXXX.XXXXXXX}
\acmConference[Conference acronym 'XX]{Make sure to enter the correct
  conference title from your rights confirmation email}{June 03--05,
  2018}{Woodstock, NY}
\acmISBN{978-1-4503-XXXX-X/2018/06}

\begin{document}

\title{UniSGR-V2: Stepwise Target-Side Feature Anticipation and Crossing for Generative Recommendation}

\author{Jun Yang}
\authornote{These authors contributed equally to this research.}
\email{yj494343@alibaba-inc.com}
\orcid{0009-0000-5546-3852}
\affiliation{%
  \institution{Alibaba International Digital
Commerce Group}
  \city{Beijing}
  \country{China}
}

\author{Jiasi Chen}
\authornotemark[1]
\email{chenaass@bupt.edu.cn}
\affiliation{%
  \institution{Alibaba International Digital
Commerce Group}
  \city{Beijing}
  \country{China}
}

\author{Jiawei Sun}
\email{xiangxing.sjw@alibaba-inc.com}
\orcid{0009-0001-6018-234X}
\affiliation{%
  \institution{Alibaba International Digital
Commerce Group}
  \city{Beijing}
  \country{China}
}

\author{Ziyue Guo}
\email{guoziyue.gzy@alibaba-inc.com}
\affiliation{%
  \institution{Alibaba International Digital
Commerce Group}
  \city{Beijing}
  \country{China}
}

\author{Jianan Yan}
\authornote{Corresponding author.}
\email{jianan.yjn@alibaba-inc.com}
\affiliation{%
  \institution{Alibaba International Digital
Commerce Group}
  \city{Beijing}
  \country{China}
}

\author{Tao Luo}
\email{luotao.lt@lazada.com}
\affiliation{%
  \institution{Alibaba International Digital
Commerce Group}
  \city{Beijing}
  \country{China}
}

% \author{Xiaoyi Zeng}
% \email{yuanhan@taobao.com}
% \affiliation{%
%   \institution{Alibaba International Digital
% Commerce Group}
%   \city{Beijing}
%   \country{China}
% }

%%
%% By default, the full list of authors will be used in the page %% headers. Often, this list is too long, and will overlap
\renewcommand{\shortauthors}{Trovato et al.}

\begin{abstract}
Semantic-ID-based generative recommendation unifies retrieval and ranking through autoregressive decoding, but it has a structural difference from discriminative models: discriminative models can directly access rich target-item side features, such as category, brand, and shop, whereas generative models cannot access these features during token-by-token decoding. The attributes are compressed into a finite codebook during quantization, while the target item itself has not yet been determined and therefore its features cannot be looked up. We refer to this state as the \textbf{target-side feature vacuum}. It simultaneously weakens code generation at every semantic-ID level, due to high conditional entropy and cascading errors across levels, and final-stage ranking, due to the absence of target-side feature crossing.

Existing methods either interact with candidate representations or autoregressively generate discrete attribute tokens, but none reconstructs, inside the decoding path, a continuous target-side representation that both conditions stepwise generation and is directly used for ranking. We propose \textbf{TacGR (Stepwise Target-Side Feature Anticipation \& Crossing for Generative Recommendation)}. Before decoding each semantic-ID code, TacGR anticipates a continuous target-side feature vector and crosses it with the semantic prefix and user behavior sequence inside the decoder. The resulting crossed representation both conditions generation at the current level and is directly used for ranking.

Regressing a continuous vector introduces two additional challenges: the risk of a trivial solution in which a learnable regression target collapses with its prediction, and the propagation of prediction noise into the generation path through attention. TacGR addresses these problems with a semantic-anchoring auxiliary tower and a unidirectional visibility mask, respectively. In addition, to address deployment challenges arising from extremely long industrial behavior histories, we introduce gated convolutional denoising to suppress noise and adapt the pyramid memory funnel to reduce inference cost. In offline experiments on a large-scale industrial dataset from Lazada, TacGR significantly outperforms recent baselines including OneRec \cite{onerecv2,onerec}, UniRec \cite{unirec}, and UniSGR \cite{unisgr}; online A/B testing further improves IPV, Order, and GMV by 0.79\%, 5.55\%, and 3.09\%, respectively.
\end{abstract}




%% ------------------------------------------------------------------
%% ACM sample teaser / received-date block retained verbatim but disabled.
%% ------------------------------------------------------------------
\iffalse
\begin{teaserfigure}
 \includegraphics[width=\textwidth]{sampleteaser}
  \caption{Seattle Mariners at Spring Training, 2010.}
  \Description{Enjoying the baseball game from the third-base
  seats. Ichiro Suzuki preparing to bat.}
  \label{fig:teaser}
\end{teaserfigure}

\received{20 February 2007}
\received[revised]{12 March 2009}
\received[accepted]{5 June 2009}
\fi

\maketitle

%% ===== TacGR paper body begins here =====
\section{Introduction}

As a key Web application, industrial recommender systems have long relied on a multi-stage cascade architecture of recall, coarse ranking, fine ranking, and reranking. Stages optimize inconsistent objectives, upstream errors propagate irreversibly downstream, and system maintenance is costly. Generative recommendation has rapidly emerged as an end-to-end alternative \cite{tiger,letter,onerec}: conditioned on the user profile and behavior history, an autoregressive decoder directly generates the identifier of the target item. TIGER \cite{tiger} first replaced random integer IDs with hierarchical semantic IDs obtained from RQ-VAE \cite{rqvae}, placing semantically similar items close in code space; since then OneRec \cite{onerecv2,onerec}, OneRanker \cite{oneranker}, RankGR \cite{rankgr}, UniRec \cite{unirec}, and UniSGR \cite{unisgr} have pushed this paradigm toward industrial scale.

Yet semantic-ID generative recommendation and discriminative models differ in a structural way that has not been directly addressed: \textbf{discriminative models can access the target item's side features throughout---structured attributes such as category, shop, and brand, which are core signals for CTR/CVR estimation---whereas generative models have no target-side features available during the entire token-by-token decoding process.} UniRec \cite{unirec} shows that the two are mathematically equivalent, and that the gap arises \textbf{not from a modeling asymmetry but from feature coverage}: structured attributes are folded into discrete codes after semantic-ID quantization. More fundamental is the \textbf{temporal contradiction}: the target item is still being generated, so its true features cannot be retrieved; although the completed semantic ID can be mapped back to an item to retrieve its features, the generation probability at every level has already been determined and cannot be corrected retrospectively. We call this structural state during decoding the \textbf{target-side feature vacuum}, which leads to two parallel consequences.

\textbf{Consequence 1: generation.} Every semantic-ID codeword is predicted without target-item attribute information. At each level, generation is conditioned only on the user context and the already decoded prefix. The first level has no prefix at all and therefore has the highest conditional entropy and is most prone to error. Because semantic IDs are hierarchical, an error at any level is irreversibly propagated to all subsequent levels.

\textbf{Consequence 2: ranking.} Without an additional post-hoc feature-lookup pipeline, scoring degenerates into ranking based only on decoder hidden states. Retrieving features by mapping the completed SID back to an item can only repair ranking and cannot benefit the decoding process itself. In most end-to-end designs, the ranking head is built directly on top of the decoder and thus lacks the target-side feature crossing on which discriminative models rely.

Most existing methods address only one of these consequences: some perform target-side interaction using candidate SID representations \cite{unisgr,rankgr,oneranker}, some map decoded SIDs back to items for post-hoc reranking \cite{gryphon}, and others adjust the generation policy through preference alignment \cite{onerecv2,onerec}---yet the model remains unaware of the target attributes throughout decoding. The work most closely aligned with our motivation is UniRec's \cite{unirec} Chain-of-Attribute (CoA): it autoregressively generates discrete attribute tokens before semantic IDs to reduce generation conditional entropy, but an incorrect attribute decision cannot be revised, and as the number of attribute dimensions grows, decoding steps increase linearly while signal bandwidth remains constrained by the vocabulary.

We propose \textbf{TacGR (Stepwise Target-Side Feature Anticipation \& Crossing for Generative Recommendation)}. TacGR progressively anticipates the target-side features missing in the vacuum inside the decoding path and crosses them with the semantic prefix and user behavior sequence inside the decoder, producing a contextualized target-side representation that serves both generation and ranking. In industrial systems, target-side features are typically continuous vectors formed by concatenating learnable embeddings; choosing to regress this vector rather than generate discrete tokens introduces two costs: the regression target lacks an independent semantic constraint, so the labels and predictions may jointly collapse to a trivial solution, and noisy predictions can propagate through attention and induce self-dependency degeneration. Around the central anticipation-and-crossing mechanism and these two derived problems, TacGR makes the following contributions.

\begin{enumerate}
\def\labelenumi{\arabic{enumi}.}
\item
  \textbf{Cascaded sideinfo anticipation (CSA) and in-decoder crossing (§3.3).} Before decoding each SID level, TacGR anticipates a continuous side-information vector and crosses it with the sequence inside the decoder. The same anticipated signal simultaneously restores target-side conditioning for generation and target-side feature crossing for ranking.
\item
  \textbf{Semantic anchoring of the regression target (§3.4).} We formally characterize the trivial-solution risk of learnable regression targets and use an auxiliary ranking tower to provide a task-aligned semantic anchor.
\item
  \textbf{Unidirectional visibility mask (UVM) (§3.5).} Semantic tokens are prevented from attending to prediction tokens, and different prediction tokens are mutually invisible. This architectural constraint blocks prediction-error propagation into the semantic hidden-state path.
\item
  \textbf{Two challenges of extremely long behavior histories (§3.7).} In large-scale industrial deployment, user histories may contain thousands of interactions, posing general challenges to generative models: noisy interactions contaminate user representations, while attention overhead squeezes the online latency budget. For the former we introduce a \textbf{gated convolutional denoising operator (GCD)}; for the latter we adapt the memory shrinking of the pyramid memory funnel \cite{onetrans} to the memory dimension of decoder cross-attention.
\end{enumerate}

We comprehensively evaluate these designs through large-scale offline experiments and online A/B testing on Lazada: TacGR outperforms UniSGR on all four retrieval metrics and all four ranking metrics, and online A/B testing further improves IPV, Order, and GMV by 0.79\%, 5.55\%, and 3.09\%, respectively.

\section{Related Work}

\subsection{Semantic IDs and Generative Recommendation}

Generative recommendation represents items as identifiers that can be generated autoregressively. DSI \cite{dsi} jointly learns document contents and IDs, TIGER \cite{tiger} constructs hierarchical semantic IDs with RQ-VAE \cite{rqvae}, and LETTER \cite{letter}, UniRec \cite{unirec}, and UniSGR \cite{unisgr} further improve code representation and assignment. Recent work unifies retrieval and ranking through candidate interaction \cite{unisgr,rankgr,oneranker}, post-hoc reranking \cite{gryphon}, or preference alignment \cite{onerecv2,onerec,onesug}, but primarily exploits candidate representations or business signals rather than providing structured target attributes during token-by-token decoding.

\subsection{Attribute-Conditioned Generation}

UniRec's \cite{unirec} Chain-of-Attribute (CoA) generates discrete attribute tokens before the semantic-ID sequence to reduce uncertainty in subsequent decoding, but its granularity is constrained by the vocabulary, and increasing the number of attribute dimensions further lengthens the decoding chain. TacGR instead anticipates a continuous target-side representation at each level and uses it for both semantic-ID generation and candidate ranking. Since true target attributes are available during training but unavailable at inference, this problem also falls within the privileged-information setting formalized by LUPI \cite{lupi}. Unlike privileged feature distillation \cite{pfd}, in which a teacher guides a student, we directly anticipate and exploit this information within autoregressive decoding.

\subsection{Extremely Long Behavior Sequences}

Long behavior sequences bring noise and computational overhead at the same time. Existing work performs denoising through interaction reweighting, attention sparsification, or frequency-domain filtering \cite{denoiseimplicit,denoisesasrec,fmlprec}, and improves efficiency through sparse attention, hierarchical compression \cite{longformer,bigbird,funnel}, MemoryNet \cite{unisgr}, Lazy Decoder \cite{onerecv2}, prefix caching \cite{davgen}, and progressive shrinking \cite{onetrans}. This work adapts gated convolutional denoising and cross-attention-oriented memory shrinking to these two classes of problems, respectively.

\section{Method}

\subsection{Problem Formulation}

Given a user \(u\), the user's behavior history \(\mathcal{H}=[h_1,\ldots,h_n]\) ordered in reverse chronological order, with the most recent interaction first, and user profile \(\mathbf{p}_u\), our goal is to generate candidates and perform multi-objective ranking. Each item \(v\) is represented by an \(L\)-level semantic ID \(\mathbf{s}_v=[s_v^1,\ldots,s_v^L]\), \(s_v^l\in\{0,\ldots,V-1\}\), with the autoregressive generation process
\begin{equation}
\begin{aligned}
P(\mathbf{s}_v\mid \mathcal{H},\mathbf{p}_u)
=
\prod_{l=1}^{L}
P\!\left(
s_v^l
\mid
\mathbf{s}_v^{<l},
\mathcal{H},
\mathbf{p}_u
\right).
\end{aligned}
\end{equation}
The ranking stage predicts three objectives for every candidate: CTR, ATC, and CVR.

Let the target item's side feature be \(\phi_\theta(v)\in\mathbb{R}^{d_\phi}\). In industrial systems, it is typically formed by concatenating learnable embeddings of several ID features, such as category, shop, and brand. The subscript \(\theta\) emphasizes that these parameters are learnable, which is crucial to the discussion in §3.4.

\textbf{Target formulation.} A discriminative ranking head can directly access \(\phi_\theta(v)\), whereas a generative model cannot obtain it during token-by-token decoding; this is precisely the target-side feature vacuum. TacGR progressively anticipates an estimate \(\hat{\phi}^{(l)}\) inside the decoding path and crosses it with the semantic prefix and user behavior sequence to obtain a contextualized target-side representation \(\tilde{\phi}^{(l)}\) (§3.3.2). The representation \(\tilde{\phi}^{(l)}\) both conditions generation at level \(l\) and directly serves ranking. Specifically, the level-\(l\) code distribution changes from \(P\!\left(s_v^l \mid \mathbf{s}_v^{<l}, \mathcal{H}, \mathbf{p}_u\right)\) to \(P\!\left(s_v^l \mid \mathbf{s}_v^{<l}, \tilde{\phi}^{(l)}, \mathcal{H}, \mathbf{p}_u\right)\), and the final ranking head changes from \(g(\mathbf{h}_{\mathrm{dec}})\) to \(g(\mathbf{h}_{\mathrm{dec}},\tilde{\phi})\).

\subsection{Overview of TacGR}

TacGR consists of five modules (Figure 1). A memory encoder transforms the behavior sequence into memory \(\mathbf{M}\in\mathbb{R}^{n\times d_m}\), and includes GCD (§3.7.1). An MoE decoder, consisting of GQA attention and shared SwiGLU-MoE blocks, autoregressively generates semantic IDs through cross-attention, whose accessible memory is scheduled by the memory funnel (§3.7.2). A cascaded sideinfo anticipator predicts \(\hat{\phi}^{(l)}\) before decoding each semantic-ID level, crosses it with the semantic prefix and behavior sequence inside the decoder, conditions the generation of semantic code \(s^l\), and simultaneously feeds the resulting representation to the ranking head (§3.3). A UVM constrains information flow within the decoder sequence (§3.5). The multitask PLE ranking head \cite{ple} takes \([\mathbf{h}_{\mathrm{dec}};\tilde{\phi}]\) as input and outputs three objectives. During training, an additional anchoring auxiliary tower is driven by the ground-truth side information (§3.4).

\begin{figure*}[t]
  \centering
  \IfFileExists{figures/tacgr_overview_new2.pdf}{
    \includegraphics[width=\textwidth]{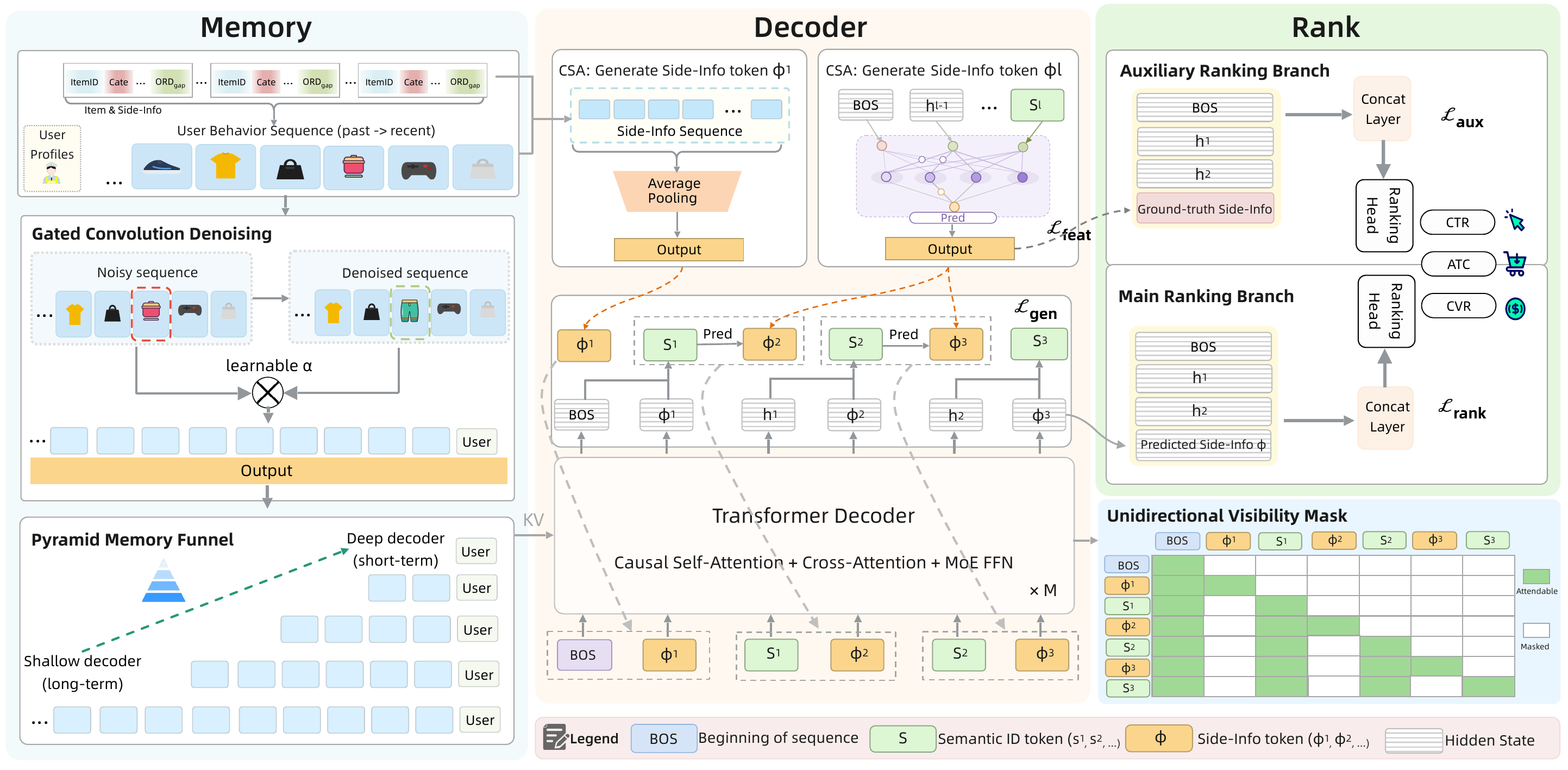}
  }{
    \fbox{\parbox[c][0.23\textheight][c]{0.96\textwidth}{\centering
    \textbf{Figure 1 placeholder}\\[4pt]
    Export the TacGR overview slide from PowerPoint as\\
    \texttt{figures/tacgr\_overview.pdf}.}}
  }
  \caption{Overall architecture of TacGR.}
  \Description{Overview of the memory encoder, decoder, stepwise target-side feature anticipation, ranking branches, memory funnel, denoising module, and visibility mask in TacGR.}
  
\end{figure*}

\subsection{Cascaded Sideinfo Anticipation and In-Decoder Crossing}

\subsubsection{Stepwise Anticipation Points and Progressively Enriched Conditions}

Anticipating side information only before the final semantic code underuses the hierarchical structure of semantic IDs. Different levels capture semantic decisions at different granularities: the first may identify a coarse category, the second narrow it to a fine-grained category, and the third further distinguish individual items. TacGR therefore places a prediction \(\hat{\phi}^{(l)}\) before every code \(s^l\), between \(s^{l-1}\) and \(s^l\), with \(s^0:=\mathrm{BOS}\). Each prediction is thus paired with the preceding semantic token, expanding the decoder input from \(L+1\) to \(2L+1\) tokens:

\begin{equation}
\begin{aligned}
[
\mathrm{BOS},
\hat{\phi}^{(1)},
s^1,
\hat{\phi}^{(2)},
s^2,
\ldots,
\hat{\phi}^{(L)},
s^L
].
\end{aligned}
\end{equation}
As decoding proceeds, each prediction point receives progressively richer conditioning information:
\begin{equation}
\hat{\phi}^{(l)}
=
\begin{cases}
\displaystyle
\frac{1}{|\mathcal{H}|}
\sum_i \phi_\theta(h_i),
& l=1,\\[6pt]
\mathrm{Pred}^{(l)}
\left(
[\mathbf{h}^{(0)};\ldots;\mathbf{h}^{(l-2)}],
e(s^{l-1})
\right),
& 2\le l\le L.
\end{cases}
\end{equation}

Here, \(\mathbf{h}^{(k)}\) is the decoder hidden state at code position \(s^k\), \(\mathbf{h}^{(0)}:=\mathbf{h}^{\mathrm{bos}}\), and \(e(\cdot)\) is the code-embedding lookup; predictors at different levels use independent parameters. For \(l\ge2\), \(\hat{\phi}^{(l)}\) is computed after \(s^{l-1}\) is sampled. The cached hidden states \(\mathbf{h}^{(0)},\ldots,\mathbf{h}^{(l-2)}\) summarize the preceding decoding context, while \(e(s^{l-1})\) directly supplies the latest level of semantic information. Together, these heterogeneous conditions allow the prediction to exploit all semantic information decoded through \(s^{l-1}\) and become progressively finer-grained. Because \(e(s^{l-1})\) is obtained by direct lookup, the design also supports the paired forward pass in §3.5.4 without an additional decoder call. Each prediction point further uses an independent token-type embedding and output projection to distinguish different prediction levels.

\textbf{Special case: the cold-start level (\(l=1\)).} The prediction \(\hat{\phi}^{(1)}\) appears before the first semantic code. With no preceding semantic code available, it is paired with BOS and conditioned entirely on a user-side prior obtained by masked mean pooling over the side information of the user's clicked history, as defined in the first branch. This prior summarizes the categories, shops, and brands in the user's recent interactions. Because it is aggregated from ground-truth side features of historical items, it is excluded from \(\mathcal{L}_{\mathrm{feat}}\). It also does not depend on decoder hidden states and is therefore compatible with paired forward inference.

\subsubsection{In-Decoder Crossing}

The anticipated vector \(\hat{\phi}^{(l)}\) from §3.3.1 is projected by \(\psi(\cdot)\) and inserted into the decoder as a token, where it is crossed with the semantic prefix and user behavior sequence. The resulting representation conditions generation of the current semantic code and supports final-stage ranking.

\textbf{How crossing occurs.} Under the UVM (§3.5), the prediction token \(\psi(\hat{\phi}^{(l)})\) reads BOS and the already decoded semantic-ID prefix through self-attention, and reads user behavior memory \(\mathbf{M}\) through cross-attention. Let the output hidden state at this prediction slot be \(\tilde{\phi}^{(l)}\), a target-side representation integrating semantic-prefix and sequence information; these attention interactions constitute \textbf{feature crossing}. The crossing is unidirectional: prediction tokens can read the semantic prefix and memory, while semantic tokens cannot attend to prediction tokens. The crossing therefore does not contaminate the semantic hidden-state path.

\textbf{For generation.} The crossed representation \(\tilde{\phi}^{(l)}\) is injected as a soft condition for generation at level \(l\):
\begin{equation}
\begin{aligned}
P(s^l\mid\cdot)
=
\mathrm{softmax}
\left(
\mathbf{W}_{s_i}^{(l)}
[
\mathbf{h}^{(l-1)};
\tilde{\phi}^{(l)}
]
\right).
\end{aligned}
\end{equation}
The generation loss can backpropagate through \(\tilde{\phi}^{(l)}\), allowing the predictor to regress side information while adapting to semantic-ID generation. The regression loss in §3.4 keeps predictions aligned with ground-truth side information, preventing unconstrained conditioning. \(\mathbf{W}_{s_i}^{(l)}\) is level-specific to match different semantic granularities.

\textbf{For ranking.} After a residual projection, the same crossed representation is concatenated with decoder hidden states and fed into the multitask PLE:
\begin{equation}
\begin{aligned}
\mathbf{x}_{\mathrm{rank}}
=
[
\mathbf{h}_{\mathrm{dec}};
\mathbf{z}_L
],
\qquad
\mathbf{z}_L
=
\mathrm{ResProj}
\left(
\tilde{\phi}^{(L)}
\right).
\end{aligned}
\end{equation}
Here, \(\mathbf{h}_{\mathrm{dec}}=[\mathbf{h}^{(0)};\mathbf{h}^{(1)};\ldots;\mathbf{h}^{(L-1)}]\) is the concatenation of the semantic hidden states produced by the decoder forward passes.

The crossed representation is not merely an intermediate product of generation but a direct ranking input, corresponding to \(\tilde{\phi}\) in §3.1.

\subsubsection{Why Continuous Vectors?}

We regress a continuous vector rather than autoregressively generate discrete attribute tokens as in CoA primarily for \textbf{inference efficiency}. Industrial systems often have over ten target-side attributes; generating them individually requires decoding steps linear in attribute count and can limit serving throughput. TacGR instead predicts a single vector \(\hat{\phi}^{(l)}\) of concatenated attribute embeddings per semantic-ID level, decoupling inference cost from attribute count. Continuous vectors also offer \textbf{error tolerance}: incorrect discrete categories irreversibly change subsequent decoding, whereas continuous prediction errors appear as slight vector displacements that smoothly perturb downstream code distributions through soft conditioning. They further offer \textbf{signal bandwidth}: unlike vocabulary-bound discrete tokens, they retain full information in the learned embedding space and express semantic gradients between two category prototypes or within relatively high-end shop segments. The trade-off is the need to address the two problems introduced in the Introduction, motivating the next two sections.

\subsection{Semantic Anchoring of the Regression Target}

\subsubsection{Trivial Solutions for Learnable Regression Targets}

A natural regression objective is
\begin{equation}
\mathcal{L}_{\mathrm{feat}}
=
\sum_{l=2}^{L}
\left\|
\hat{\phi}^{(l)}-\phi_\theta(v)
\right\|_2^2.
\end{equation}
Here, \(\hat{\phi}^{(l)}\) is the target-side information predicted at level \(l\), while the label \(\phi_\theta(v)\) is formed by concatenating learnable ID embeddings for the target item's category, shop, brand, and other features. Unlike a fixed supervision target, \(\phi_\theta(v)\) changes as \(\theta\) is jointly updated with the predictor, allowing the prediction and its regression target to move together and thereby admitting degenerate solutions.

\textbf{Proposition 1 (Degeneracy of the regression target).} If \(\theta\) receives gradients only through \(\mathcal{L}_{\mathrm{feat}}\), there exists a parameter setting such that \(\mathcal{L}_{\mathrm{feat}}=0\) and \(I(\phi_\theta(v);v)=0\).

\textbf{Proof.} Let \(\phi_\theta(\cdot)\equiv\mathbf{c}\) and let the predictor always output \(\mathbf{c}\). All regression residuals are then zero, while \(\phi_\theta(v)\) is constant and thus has zero mutual information with \(v\).

Therefore, convergence of the regression loss does not guarantee an informative side representation. Even without exact constant collapse, the representation may drift toward a low-rank, low-variance solution \cite{collapse} and lose item-discriminative information. In our system, the category, shop, brand, and other feature embeddings that form the target-side label are also used to construct historical-item representations, and therefore receive gradients from the generation and ranking objectives through behavior-sequence modeling. These gradients discourage exact constant collapse but do not directly constrain \(\phi_\theta(v)\) in its role as a target label: the main ranking path consumes representations derived from the prediction, whereas the embeddings forming the ground-truth label receive indirect discriminative supervision through sequence aggregation when their attribute values occur in the history. Because the target-item and historical-item distributions do not fully overlap, particularly for tail items, this indirect constraint remains insufficient. The anchoring tower in §3.4.2 therefore connects the ground-truth target-side representation directly to the business objectives; §4.3 provides the corresponding ablation evidence.

\subsubsection{Anchoring with an Auxiliary Ranking Tower}

We introduce an anchoring auxiliary tower that exists only during training. Instead of taking the predicted vector as input, it consumes the true side information \(\phi_\theta(v)\), concatenates it with the shared decoder hidden state, and passes the result through an independent PLE and three independent decision heads. The BCE objective over the three business targets \(\mathcal{T}=\{\mathrm{ctr},\mathrm{atc},\mathrm{cvr}\}\) uses the same supervision conditions as the main ranking head (§3.6):
\begin{equation}
\begin{aligned}
\mathcal{L}_{\mathrm{anchor}}
&= \sum_{t\in\mathcal{T}}
\lambda_t\,
\mathrm{BCE}\bigl(\hat{y}^{\mathrm{aux}}_t,\, y_t\bigr), \\
\hat{y}^{\mathrm{aux}}_t
&= \mathrm{Head}^{\mathrm{aux}}_t
\Bigl(
\mathrm{PLE}^{\mathrm{aux}}
\bigl(
[\mathbf{h}_{\mathrm{dec}};\ \mathrm{ResProj}(\phi_\theta(v))]
\bigr)
\Bigr).
\end{aligned}
\end{equation}

This loss supplies \(\theta\) with gradients aligned to the business objectives. If \(\phi_\theta\) collapses to a constant, the auxiliary tower degenerates into a predictor that depends only on \(\mathbf{h}_{\mathrm{dec}}\), and \(\mathcal{L}_{\mathrm{anchor}}\) cannot decrease below the level achievable by the shared backbone alone. Collapse is therefore discouraged. Intuitively, the auxiliary tower encourages \(\phi_\theta\) to converge to a semantic space that remains discriminative for click, add-to-cart, and pay. Only then is it meaningful for \(\hat{\phi}\) to approximate that space: the predictor learns signals such as the likely category of the item, the tier of its shop, and its conversion tendency, rather than an uninformative vector.

The auxiliary tower is executed only during training and is completely removed at inference time, so it introduces no online serving overhead.

\subsection{Unidirectional Visibility Mask}

\subsubsection{Risk: Error Propagation and Self-Dependency Degeneration}

Treating the predicted vector \(\hat{\phi}^{(l)}\) as an ordinary token introduces a non-negligible risk.

Under a standard grouped causal mask, the level-\(l\) semantic code \(s^l\) can attend to all preceding prediction tokens \(\hat{\phi}^{(1)},\ldots,\hat{\phi}^{(l)}\), and prediction tokens can also attend to one another. Noisy signals can therefore propagate along multiple paths in the attention graph. A more subtle problem is that the predictor and generator reside in the same computation graph: the generator may learn to accommodate systematic predictor errors. Gradients from the generation loss can further reinforce this coupling through the prediction branch, eventually creating a self-consistent internal loop that drifts away from the true data distribution.

\subsubsection{Mask Definition}

We organize decoder inputs into item groups, with each group containing \(2L\) tokens corresponding to one candidate item. Different groups are mutually invisible and share only a globally visible BOS token. This exactly matches inference, where each beam candidate independently expands from BOS.

Let the set of relative positions within a group be \(\{0,\ldots,G-1\}\), with prediction-token positions denoted by \(\mathcal{P}\) and semantic-token positions by \(\mathcal{S}\). We define the UVM \(\mathbf{A}\) such that query position \(i\) can attend to key position \(j\) if and only if
\begin{equation}
j=\mathrm{BOS}
\;\lor\;
j=i
\;\lor\;
(j<i\land j\in\mathcal{S}).
\end{equation}

In other words, BOS is globally visible, each token can attend to itself, and a token may look backward only at semantic tokens. Except for a prediction token attending to itself, no position is allowed to attend to prediction tokens. In particular, different prediction tokens are mutually invisible.

For \(L=3\), consider the six-token layout \([\hat{\phi}^{(1)}, s^1, \hat{\phi}^{(2)}, s^2, \hat{\phi}^{(3)}, s^3]\), with \(\mathcal{P}=\{0,2,4\}\), \(\mathcal{S}=\{1,3,5\}\). Then \(s^2\) at position 3 can attend only to BOS, \(s^1\) at position 1, and itself. The prediction token \(\hat{\phi}^{(3)}\) at position 4 can attend to BOS, \(s^1\), \(s^2\), and itself, but cannot attend to \(\hat{\phi}^{(2)}\).

\subsubsection{Error-Isolation Analysis}

Under teacher forcing, define the level-\(l\) prediction error as \(\epsilon^{(l)} = \hat{\phi}^{(l)} - \phi_\theta(v)\). Under a standard causal mask, \(s^l\) can attend to \(\hat{\phi}^{(1)},\ldots,\hat{\phi}^{(l)}\), so the semantic hidden state \(\mathbf{h}^{(l)}\) becomes a function of \(\epsilon^{(1)},\ldots,\epsilon^{(l)}\): larger prediction errors induce larger generation deviations, and the errors accumulate across levels. The UVM completely breaks this coupling. Semantic tokens cannot attend to any prediction token, so none of the prediction errors \(\epsilon^{(1)},\ldots,\epsilon^{(l)}\) can propagate into \(\mathbf{h}^{(l)}\) through attention. The semantic hidden-state path therefore behaves as if the prediction tokens did not exist.

Now consider the prediction side. For \(l\ge2\), the conditioning inputs of \(\hat{\phi}^{(l)}\) consist of two parts: hidden states \(\mathbf{h}^{(0)},\ldots,\mathbf{h}^{(l-2)}\) and the code embedding \(e(s^{l-1})\). As established above, prediction errors cannot propagate into any \(\mathbf{h}^{(k)}\) through attention. Under teacher forcing, \(s^{l-1}\) is the ground-truth code, so \(e(s^{l-1})\) also carries no prediction error. Every prediction level therefore starts from a clean base. Even if the level-\((l-1)\) prediction is inaccurate, the error cannot contaminate the level-\(l\) conditioning input through either hidden states or semantic-code embeddings. Prediction errors are isolated level by level and cannot snowball through the hierarchy.

\subsubsection{Inference Optimization: Paired Forward Passes}

A naive autoregressive implementation processes \(L\) semantic tokens and \(L\) prediction tokens in separate decoder calls, requiring \(2L\) calls. Because the level-\(l\) predictor depends only on cached preceding semantic hidden states and the code embedding \(e(s^{l-1})\), \(\hat{\phi}^{(l)}\) can be computed immediately after \(s^{l-1}\) is sampled. We therefore pair \(s^{l-1}\) and \(\psi(\hat{\phi}^{(l)})\) as a two-token input in a single decoder call. A composite attention mask allows the former to access the cached semantic prefix and itself, while allowing the latter to additionally access \(s^{l-1}\) from the same call. This jointly produces \(\mathbf{h}^{(l-1)}\) for level-\(l\) code generation and \(\tilde{\phi}^{(l)}\) for generation and ranking, reducing the number of sequential decoder calls from \(2L\) to \(L\) without introducing additional calls relative to the baseline. In contrast, UniRec (CoA) decodes attribute tokens sequentially because of inter-attribute dependencies, lengthening generation and preventing paired execution. Appendix~\ref{app:efficiency} reports serving benchmarks for full TacGR, UniRec (CoA), and TacGR without paired forward.

\subsection{Training Objective}

The total objective is a weighted sum of four losses:

\begin{equation}
\begin{aligned}
\mathcal{L}
&= \lambda_{\mathrm{gr}}\mathcal{L}_{\mathrm{GR}}
+ \lambda_{\mathrm{rank}}\mathcal{L}_{\mathrm{rank}} \\
&\quad + \lambda_{\mathrm{feat}}\mathcal{L}_{\mathrm{feat}}
+ \lambda_{\mathrm{anchor}}\mathcal{L}_{\mathrm{anchor}}.
\end{aligned}
\end{equation}
The generation loss \(\mathcal{L}_{\mathrm{GR}}\) is a behavior-weighted cross-entropy over semantic IDs. For each \((\text{user},\text{target item})\) sample \((u,v)\), the \(L\) semantic codes are supervised against their ground-truth values \(s_\star^l\), and the entire sample is weighted according to its behavior type:
\begin{equation}
\begin{aligned}
\mathcal{L}_{\mathrm{GR}}
=
-
\sum_{(u,v)}
w_{b(u,v)}
\sum_{l=1}^{L}
\log
P
\left(
s^l=s_\star^l
\mid
\mathbf{s}^{<l},
\tilde{\phi}^{(l)},
\mathbf{M}
\right).
\end{aligned}
\end{equation}
Here, \(b(u,v) \in \{\text{impression}, \text{click}, \text{add-to-cart}, \text{pay}\}\) is the highest-value behavior type associated with the sample, and \(w_b\) is its weight, satisfying \(w_{\mathrm{pv}} < w_{\mathrm{click}} < w_{\mathrm{add\text{-}to\text{-}cart}} < w_{\mathrm{pay}}\), which biases the generation distribution toward higher-value behaviors. The level-wise conditional probability \(P(s^l\mid\cdot)\) is the soft-conditioned generation distribution defined in §3.3.2, and \(\mathbf{M}\) is the user's long-term behavior memory (§3.7).

The ranking loss \(\mathcal{L}_{\mathrm{rank}}\) is a multitask BCE over CTR, ATC, and CVR. The ATC and CVR objectives are conditioned on clicks, with their losses backpropagated only for clicked samples. Let \(\ell_t=\mathrm{BCE}(\hat{y}_t,y_t)\). Then

\begin{equation}
\mathcal{L}_{\mathrm{rank}}
=
\sum_{(u,v)}
\left[\ell_{\mathrm{ctr}}+y_{\mathrm{click}}(\ell_{\mathrm{atc}}+\ell_{\mathrm{cvr}})\right].
\end{equation}

Here, \(y_{\mathrm{click}}\in\{0,1\}\) is the click indicator, and \(y_t\) is the ground-truth label for task \(t\). The three decision heads consume the ranking input \(\mathbf{x}_{\mathrm{rank}}\) defined in §3.3.2, allowing \(\tilde{\phi}^{(L)}\) to participate directly in ranking and receive discriminative gradients. \(\mathcal{L}_{\mathrm{feat}}\) and \(\mathcal{L}_{\mathrm{anchor}}\) are defined in §3.4.1 and §3.4.2, respectively.

\subsection{Two Challenges of Extremely Long Behavior Histories}

The mechanisms in §3.3--§3.5 are all built on a long-term memory \(\mathbf{M}\): cross-attention reads user interests from it, and every side-information prediction point depends on it upstream. In industrial deployment, both memory quality and inference cost directly determine whether these mechanisms remain effective with extremely long histories. First, \textbf{noise degrades memory quality}: misclicks, impressions that users immediately scroll past, and other noisy interactions perturb user-interest representations and affect both the generation path and the side-information regression path. Second, \textbf{online latency budgets limit memory length}: every decoder layer repeatedly scans the complete memory. This section presents separate designs for the noise and efficiency constraints.

\subsubsection{Noise: Gated Convolutional Denoising}

The denoising module is embedded in the memory encoder (Figure 2). Operating on the longest sequences in every forward pass, it must keep parameter count and computational complexity nearly negligible without mistaking informative interest transitions for noise. We apply depthwise convolution for temporal smoothing only once during memory encoding, without repetition across beams or decoder layers. This avoids the sequence-length-dependent transformation cost of frequency-domain filtering and irreversible information loss from mistakenly removing interactions with hard denoising.

\begin{figure}[t]
  \centering
  \IfFileExists{figures/gated_conv_denoising_new3.pdf}{
    \includegraphics[width=\columnwidth]{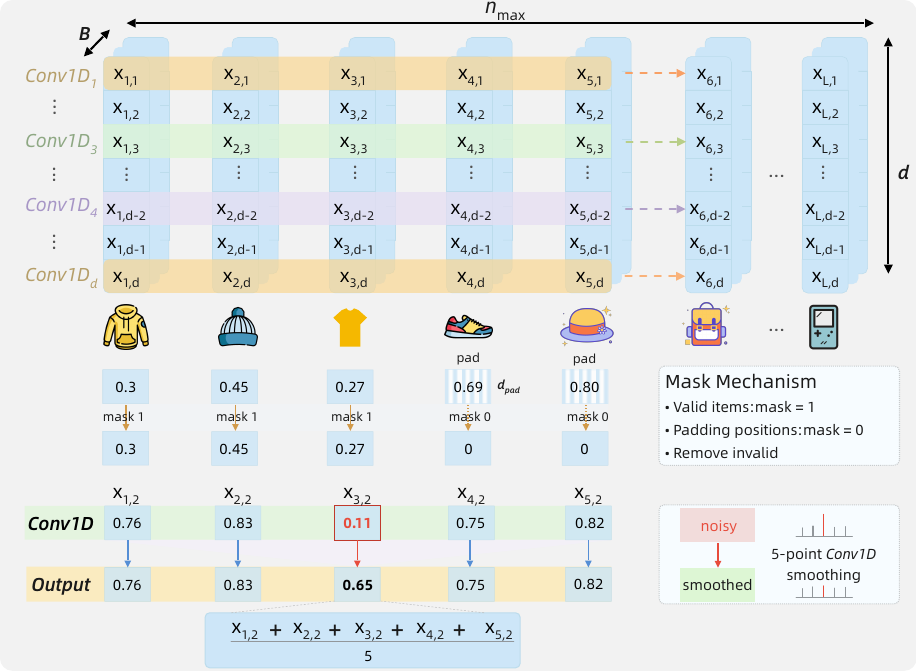}
  }{
    \fbox{\parbox[c][0.18\textheight][c]{0.92\columnwidth}{\centering
    \textbf{Figure 2 placeholder}\\[4pt]
    Export the gated convolutional denoising slide from PowerPoint as\\
    \texttt{figures/gated\_conv\_denoising.pdf}.}}
  }
  \caption{Gated convolutional denoising module.}
  \Description{The module smooths the masked behavior sequence with depthwise convolution and combines the smoothed and original views through a gated residual before memory projection.}
  
\end{figure}

Let the padded batch sequence length be \(n_{\max}\) and the position-encoded behavior sequence be \(\mathbf{x} \in \mathbb{R}^{B\times n_{\max}\times d}\). The valid-position mask \(\mathbf{m} \in \{0,1\}^{B\times n_{\max}}\) is 1 for real interactions and 0 for padding. To prevent padding-induced convolutional bias, we first zero padding (\(\mathbf{x}'=\mathbf{x}\odot\mathbf{m}\)), then apply a depthwise 1D convolution (kernel width \(k\), same padding) to each channel along time, yielding a locally smoothed estimate \(\mathbf{s}=\mathrm{Conv1D}(\mathbf{x}')\). Initialized as \(w_i=1/k\), the kernel learns non-uniform weights adapted to the noise distribution through end-to-end training. A learnable scalar \(\alpha=\sigma(g)\) controls smoothing strength through a gated residual that linearly mixes the original and smoothed signals:
\begin{equation}
\tilde{\mathbf{x}}
=
(1-\alpha)\mathbf{x}'
+
\alpha\mathbf{s}.
\end{equation}

The denoised view \(\tilde{\mathbf{x}}\) does not directly replace the original representation. Instead, it is concatenated along the feature dimension with the original, unmasked view \(\mathbf{x}\), and the concatenation is linearly projected to form long-term memory \(\mathbf{M}\), which is read by decoder cross-attention:
\begin{equation}
\begin{aligned}
\mathbf{M}
=
[
\mathbf{x};
\tilde{\mathbf{x}}
]
\mathbf{W}_M,
\qquad
\mathbf{W}_M
\in
\mathbb{R}^{2d\times d_m}.
\end{aligned}
\end{equation}
The concatenated \(\mathbf{x}\) preserves the complete representation, ensuring that the original signal participates in downstream projection without information loss. The valid-position mask \(\mathbf{m}\) used for denoising and the padding mask used by downstream cross-attention are derived from the same definition, \(\mathrm{pad\_mask} = \neg\mathbf{m}\), which guarantees that the denoising and attention stages judge valid positions identically.

\subsubsection{Efficiency: Pyramid Memory Funnel}

Every decoder layer performs cross-attention over the complete memory, with beam-search cost \(O(D\cdot K\cdot n\cdot d_m)\), making long sequences difficult to serve within an online latency budget. Cross-attention analysis of a trained model shows that, with increasing depth, semantic-ID queries attend more to user-profile information and recent history, while distant positions receive less attention.

We therefore borrow the pyramidal idea from OneTrans \cite{onetrans} to progressively shrink the visible memory. Unlike OneTrans, which shrinks the query set in single-tower self-attention, our encoder-decoder queries (semantic-ID tokens) are short and fixed in length, so we shrink memory (keys/values):

We define a decreasing memory schedule \(\boldsymbol{\tau} = [\tau_1, \ldots, \tau_D]\). Cross-attention at layer \(i\) reads only the first \(\tau_i\) memory positions, i.e., \(\mathbf{M}^{(i)}=\mathbf{M}[:,:\tau_i,:]\).

The behavior sequence is ordered reverse chronologically, so retaining a prefix preserves recent interactions, consistent with the attention analysis. This is an engineering adaptation for encoder-decoder architectures, not an independent methodological innovation, allowing §3.3--§3.5 to meet latency budgets on extremely long histories.

\section{Experiments}

We organize the experiments around five research questions:

\begin{itemize}
\item
  \textbf{RQ1:} How does TacGR compare with existing generative recommendation baselines in overall retrieval and ranking performance?
\item
  \textbf{RQ2:} What are the respective contributions of CSA, in-decoder crossing, and the stepwise design?
\item
  \textbf{RQ3:} What are the respective roles of the anchoring auxiliary tower and the UVM, and are the two complementary?
\item
  \textbf{RQ4:} Is noise in extremely long behavior histories effectively handled?
\item
  \textbf{RQ5:} How does TacGR compare with a post-hoc true-feature lookup approach built on the same backbone?
\end{itemize}

\subsection{Experimental Setup}

\textbf{Dataset.} We build an offline dataset from Lazada recommendation logs, covering four user behaviors---PV, CLK, ATC, and PAY---and four target-side attributes used as side information for anticipation and crossing---L1 category, leaf category, brand, and shop. We split the logs chronologically, using the first 30 days for training and the following day for testing. The training set contains 2.42 billion samples, covering 6.46 million users and 7.53 million items.

\textbf{Baselines.} We compare against TIGER \cite{tiger}, OneRec-V1 \cite{onerec}, OneRec-V2 \cite{onerecv2}, UniRec (CoA) \cite{unirec}, and UniSGR \cite{unisgr} as generative recommendation baselines. UniRec (CoA) is the most direct comparison to our method; we reproduce its attribute-chain mechanism on the same backbone to eliminate backbone differences. UniSGR is included as a unified generation-and-ranking framework for a more direct comparison of ranking performance.

\textbf{Metrics.} On the retrieval side, we report HitRate@100 and HitRate@500 for click (CLK) and pay (PAY) objectives. On the ranking side, we report AUC and GAUC for click and pay.

\textbf{Implementation.} TacGR uses an 8-layer decoder with an attention dimension of 512, grouped-query attention (GQA; every four query heads share one KV head), and SwiGLU-MoE with 64 experts and top-2 routing. Both the embedding dimension and the side-information dimension are 128. The side information is formed by concatenating four 32-dimensional embeddings for leaf category, L1 category, shop, and brand. We use three-level semantic IDs, with a codebook size of 2,048 at each level. The maximum behavior-sequence length is 1024.

\subsection{Main Results (RQ1)}

\begin{table*}[t]
\caption{Main retrieval and ranking results averaged over five runs with different random seeds.}\centering
\resizebox{\textwidth}{!}{%
\begin{tabular}{lrrrrrrrr}
\toprule
Method & CLK HR@100 & PAY HR@100 & CLK HR@500 & PAY HR@500 & CLK AUC & PAY AUC & CLK GAUC & PAY GAUC \\
\midrule
TIGER~\cite{tiger} & 0.2839 & 0.3420 & 0.5296 & 0.5887 & N.A. & N.A. & N.A. & N.A. \\
OneRec-V1~\cite{onerec} & 0.2980 & 0.3679 & 0.5567 & 0.6200 & N.A. & N.A. & N.A. & N.A. \\
OneRec-V2~\cite{onerecv2} & 0.2931 & 0.3570 & 0.5485 & 0.6006 & N.A. & N.A. & N.A. & N.A. \\
UniRec (CoA)~\cite{unirec} & 0.3095 & 0.3812 & 0.5739 & 0.6392 & N.A. & N.A. & N.A. & N.A. \\
UniSGR~\cite{unisgr} & 0.3001 & 0.3779 & 0.5636 & 0.6310 & 0.6752 & 0.8440 & 0.6225 & 0.5991 \\
\textbf{TacGR (Ours)} & \textbf{0.3189} & \textbf{0.3898} & \textbf{0.5861} & \textbf{0.6499} & \textbf{0.6809} & \textbf{0.8506} & \textbf{0.6292} & \textbf{0.6048} \\
\bottomrule
\end{tabular}}
\end{table*}

\textbf{Analysis.} As shown in Table 1, TacGR achieves the best results on both HR@100 and HR@500 for both click and pay behaviors. Compared with UniSGR, TacGR improves all four retrieval metrics, with a maximum relative gain of 6.26\%. Compared with UniRec (CoA) reproduced on the same backbone, the relative improvement across the four metrics ranges from 1.67\% to 3.04\%, indicating that stepwise continuous target-side feature anticipation provides effective target-side conditioning for semantic-ID generation.

Table 1 further shows that, relative to UniSGR, TacGR improves AUC by +0.84\% and +0.78\% on click and pay tasks, respectively, and improves GAUC by +1.08\% and +0.95\%, respectively. These results indicate that the target-side representation obtained through anticipation and crossing also improves downstream ranking.

\subsection{Ablation Study}

To answer RQ2--RQ4, we conduct ablations on CSA and crossing, the anchoring auxiliary tower and UVM, and GCD. We additionally design a joint ablation to test whether the anchoring auxiliary tower and UVM are complementary.

\begin{table*}[t]
\caption{Ablation results on retrieval and ranking metrics; TacGR (Full) is the shared reference.}\centering
\resizebox{\textwidth}{!}{%
\begin{tabular}{lrrrrrrrr}
\toprule
Variant & CLK HR@100 & PAY HR@100 & CLK HR@500 & PAY HR@500 & CLK AUC & PAY AUC & CLK GAUC & PAY GAUC \\
\midrule
TacGR (Full) & \textbf{0.3189} & \textbf{0.3898} & \textbf{0.5861} & \textbf{0.6499} & \textbf{0.6809} & \textbf{0.8506} & \textbf{0.6292} & \textbf{0.6048} \\
\midrule
\multicolumn{9}{l}{\textit{(a) Cascaded Sideinfo Anticipation \& In-Decoder Crossing}} \\
\quad w/o CSA \& Crossing & 0.3072 & 0.3796 & 0.5716 & 0.6360 & 0.6767 & 0.8463 & 0.6238 & 0.5994 \\
\quad w/o Crossing & 0.3162 & 0.3823 & 0.5773 & 0.6422 & 0.6788 & 0.8496 & 0.6275 & 0.6028 \\
\quad w/o Cold-Start Stage & 0.3162 & 0.3851 & 0.5769 & 0.6477 & 0.6792 & 0.8484 & 0.6272 & 0.6030 \\
\quad Last Stage Only & 0.3131 & 0.3814 & 0.5753 & 0.6376 & 0.6792 & 0.8494 & 0.6277 & 0.6040 \\
\quad w/o Anchoring Auxiliary Tower & 0.3126 & 0.3818 & 0.5770 & 0.6442 & 0.6798 & 0.8496 & 0.6272 & 0.6041 \\
\quad w/o UVM & 0.3127 & 0.3711 & 0.5736 & 0.6346 & 0.6801 & 0.8502 & 0.6283 & 0.6035 \\
\quad w/o UVM + w/o Anchoring Tower & 0.3100 & 0.3794 & 0.5725 & 0.6396 & 0.6788 & 0.8488 & 0.6258 & 0.6028 \\
\midrule
\multicolumn{9}{l}{\textit{(b) Gated Convolutional Denoising}} \\
\quad w/o GCD & 0.3110 & 0.3856 & 0.5773 & 0.6343 & 0.6797 & 0.8486 & 0.6273 & 0.6034 \\
\bottomrule
\end{tabular}}
\end{table*}

\subsubsection{Key Findings}

\textbf{CSA and crossing (RQ2).} Removing CSA and crossing degrades retrieval and ranking, with CLK HR@100 and AUC dropping by 3.67\% and 0.62\%, respectively. Retaining only final-stage anticipation or removing the cold-start stage also reduces retrieval performance, indicating that CSA and the first-stage history prior continuously supply target-side information matched to the current semantic granularity. Removing crossing alone likewise degrades both tasks, validating level-wise in-decoder crossing of anticipated features. For all four attributes, cosine similarity between predicted and ground-truth side information increases with richer semantic prefixes, consistent with the stepwise anticipation ablations; see Appendix~\ref{app:reconstruction}.

\textbf{Anchoring auxiliary tower and UVM (RQ3).} Both modules contribute to retrieval and ranking: removing the tower decreases CLK HR@100 and GAUC by 1.98\% and 0.32\%, respectively, while removing the UVM decreases them by 1.94\% and 0.14\%. Removing both yields larger drops of 2.79\% and 0.54\%, respectively, and more pronounced degradation on multiple retrieval metrics, indicating some complementarity: the tower provides task-semantic constraints, while the UVM isolates prediction-error propagation and self-dependency paths.

\textbf{Extremely long sequence processing (RQ4).} Removing GCD lowers CLK HR@100 and AUC by 2.48\% and 0.18\%, respectively, supporting its role in reducing noise interference in extremely long histories.

\subsection{Post-Hoc True-Feature Lookup (RQ5)}

TacGR and the post-hoc lookup variant share the same backbone and GCD. TacGR adds CSA with in-decoder crossing, the anchoring auxiliary tower, and the UVM; the variant instead feeds true target-side features to the ranking head only after SID-to-item mapping, so these features cannot affect candidate generation. 

\begin{table*}[t]
\caption{Comparison with post-hoc true-feature lookup. Each HitRate entry reports HR@100/HR@500. Gen. uses generation scores; Fused multiplies generation scores by CTR for CLK and by CTR × CVR for PAY.}
\label{tab:true_feature_lookup}
\centering
\resizebox{\textwidth}{!}{%
\begin{tabular}{lcccccc}
\toprule
& \multicolumn{2}{c}{Ranking}
& \multicolumn{2}{c}{CLK HitRate}
& \multicolumn{2}{c}{PAY HitRate} \\
\cmidrule(lr){2-3}
\cmidrule(lr){4-5}
\cmidrule(lr){6-7}
Method
& CLK AUC/GAUC
& PAY AUC/GAUC
& Gen.
& Fused
& Gen.
& Fused \\
\midrule
True-Feature Lookup
& \textbf{0.6854/0.6359}
& \textbf{0.8515/0.6107}
& 0.3028/0.5672
& 0.3272/0.5962
& 0.3753/0.6315
& 0.4241/0.6632 \\
TacGR (Ours)
& 0.6809/0.6292
& 0.8506/0.6048
& \textbf{0.3189/0.5861}
& \textbf{0.3352/0.6059}
& \textbf{0.3898/0.6499}
& \textbf{0.4319/0.6768} \\
\bottomrule
\end{tabular}}
\end{table*}

Table~\ref{tab:true_feature_lookup} shows higher AUC/GAUC for true-feature lookup but higher generation and fused HitRate for TacGR. Post-hoc reranking cannot recover relevant items omitted during generation, despite higher AUC/GAUC. Relative to the shared backbone with GCD (w/o CSA \& Crossing in Table 2), TacGR recovers about 46\% of the GAUC gain from lookup. The tested fusion further improves both methods' HitRate and narrows their gap by 26--50\%, highlighting the ranking head's contribution to candidate reranking without eliminating the gap inherited from generation. Additionally, deploying the post-hoc lookup approach requires multiple online services for generation, lookup, and ranking, whereas deploying TacGR requires only one. Appendix~\ref{app:efficiency} reports the corresponding serving benchmarks.

\subsection{Online A/B Test}

\begin{table}[t]
\caption{Online A/B testing results (all improvements are statistically significant with \(p<0.05\)).}\centering
\begin{tabular}{lrrr}
\toprule
Scenario & IPV & Order & GMV \\
\midrule
Lazada homepage Guess-You-Like & +0.79\% & +5.55\% & +3.09\% \\
\bottomrule
\end{tabular}
\end{table}

We deploy TacGR in the Lazada homepage Guess-You-Like scenario for a 40-day online A/B test. The experiment group covers 10\% of online traffic, replacing only the UniSGR~\cite{unisgr} model that serves as one retrieval channel with TacGR. Within this channel, the unchanged score-fusion rule combines generation and ranking scores, so its candidates require no additional pre-ranking stage before being merged with pre-ranked candidates from other retrieval channels and sent to the downstream ranking stage; all other online configurations remain unchanged. TacGR improves IPV, Order, and GMV by 0.79\%, 5.55\%, and 3.09\% over UniSGR, respectively; all gains are statistically significant.

\section{Conclusion}

We propose TacGR, which progressively anticipates and crosses continuous target-side features while the target item has not yet been determined, jointly improving semantic-ID generation and candidate ranking. The semantic-anchoring auxiliary tower and the UVM respectively constrain the regression target and isolate prediction errors. On the Lazada industrial dataset, TacGR outperforms UniSGR on both offline retrieval and ranking. In online A/B testing, TacGR improves IPV, Order, and GMV by 0.79\%, 5.55\%, and 3.09\%, respectively.

% \clearpage
\FloatBarrier
\bibliographystyle{ACM-Reference-Format}
\bibliography{references}

%%
%% Optional appendix. Add reproducibility statements, proofs, pseudocode,
%% or additional experiments here later.
\clearpage
\appendix
% \section{Additional Reproducibility Details}

\section{Side-Information Reconstruction Quality Analysis}
\label{app:reconstruction}

\begin{figure}[htbp]
% \begin{figure}[H]
  \centering
  \IfFileExists{figures/reconstruction_quality.pdf}{
    \includegraphics[width=\columnwidth]{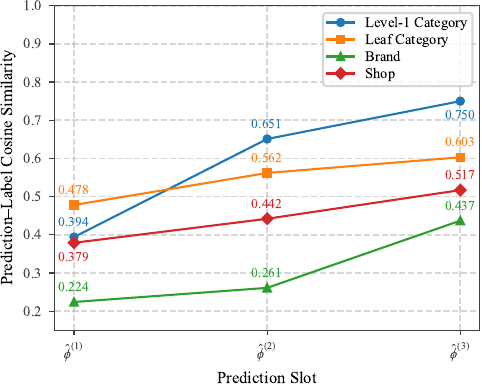}
  }{
    \fbox{\parbox[c][0.17\textheight][c]{0.92\columnwidth}{\centering
    \textbf{Figure 3 placeholder}\\[4pt]
    Side-information reconstruction quality across prediction slots and target attributes.}}
  }
  \caption{Comparison of side-information reconstruction quality across prediction slots and target attributes.}
  \label{fig:reconstruction}
  \Description{Cosine similarity between predicted and ground-truth side-information components across progressively conditioned prediction slots.}
  
\end{figure}

To further verify whether the anticipation module effectively reconstructs target-side features, we examine the outputs of the three prediction slots: \(\hat{\phi}^{(1)}\), the cold-start prediction based on the mean side information of historical items; \(\hat{\phi}^{(2)}\), conditioned on \(s^1\); and \(\hat{\phi}^{(3)}\), conditioned on \(s^1\) and \(s^2\). We split each predicted vector according to the four 32-dimensional ID-feature components of \(\phi_\theta\)---leaf category, L1 category, shop, and brand---and compute cosine similarity between each component and its ground-truth counterpart. The results are shown in Figure~\ref{fig:reconstruction}.

All four attributes exhibit the monotonic pattern \(\hat{\phi}^{(1)} < \hat{\phi}^{(2)} < \hat{\phi}^{(3)}\), demonstrating the consistent benefit of progressively richer conditioning for side-information reconstruction. However, the magnitude of improvement differs across attributes. Reconstruction quality for the L1 category nearly doubles from cold start to full conditioning; leaf category and shop lie in the middle, while brand improves more moderately and remains the lowest among the four attributes throughout.

We attribute this difference to category cardinality and the geometry of the embedding spaces. The L1 category has the smallest number of values, yielding a relatively sparse and uniform embedding space in which the predictor has sufficient capacity to learn a reliable mapping for every class; its reconstruction quality therefore nearly doubles. In contrast, brands are numerous and exhibit a pronounced head-heavy frequency distribution: the top 5\% of brands account for 83.1\% of the training samples. The resulting embedding space is dense and difficult to separate. Under this constraint, a finite-capacity regression model has difficulty extracting precise brand-level distinctions. Brand reconstruction quality therefore improves consistently but remains lower in absolute terms than the other attributes.

% \newpage

Overall, the benefit of stepwise conditioning follows a clear hierarchical pattern: each additional decoded semantic code provides one additional level of precision to the prediction. This monotonic improvement corroborates the downstream gains of stepwise anticipation observed in the §4.3 ablation study. The reconstruction analysis shows that \(\hat{\phi}\) becomes progressively more accurate as conditioning information increases, while the ablation study confirms that this improvement translates into actual gains in downstream retrieval and ranking.

\FloatBarrier
\section{Serving Efficiency}
\label{app:efficiency}

We benchmark the six representative configurations in Table~\ref{tab:serving_efficiency} in the same production serving environment, with identical hardware allocation and request load. UniSGR is the production baseline described in §4.5. We report throughput in QPS, average latency, and P99 latency in milliseconds.

TacGR improves throughput by 46.0\% over UniRec while reducing average and P99 latency by 0.5 and 1.4 ms, respectively. Paired forward improves throughput by 30.8\% over the unpaired implementation and reduces average and P99 latency by 2.5 and 5.4 ms, respectively. Compared with post-hoc true-feature lookup, TacGR improves throughput by 7.7\% while reducing average and P99 latency by 34.6 and 29.3 ms, respectively. Compared with UniSGR and the last-stage-only variant, full TacGR has lower throughput but remains within the deployment latency requirement; its effectiveness gains are demonstrated by the online A/B test and the ablation results, respectively.

\begin{table}[H]
\caption{Serving efficiency of six representative configurations under identical settings.}
\label{tab:serving_efficiency}
\centering
\resizebox{\columnwidth}{!}{%
\begin{tabular}{lrrr}
\toprule
Method & Throughput (QPS) $\uparrow$ & Avg. Lat. (ms) $\downarrow$ & P99 Lat. (ms) $\downarrow$ \\
\midrule
UniSGR (Online Baseline) & 1133 & 75.3 & 83.5 \\
UniRec (CoA) & 480 & 76.1 & 85.2 \\
True-Feature Lookup & 651 & 110.2 & 113.1 \\
TacGR w/o Paired Forward & 536 & 78.1 & 89.2 \\
TacGR, Last Stage Only & 1010 & 75.4 & 83.6 \\
TacGR (Full) & 701 & 75.6 & 83.8 \\
\bottomrule
\end{tabular}}
\end{table}

\iffalse
% Original ACM sample appendix retained verbatim for reference:
\appendix

\section{Research Methods}

\subsection{Part One}

Lorem ipsum dolor sit amet, consectetur adipiscing elit. Morbi
malesuada, quam in pulvinar varius, metus nunc fermentum urna, id
sollicitudin purus odio sit amet enim. Aliquam ullamcorper eu ipsum
vel mollis. Curabitur quis dictum nisl. Phasellus vel semper risus, et
lacinia dolor. Integer ultricies commodo sem nec semper.

\subsection{Part Two}

Etiam commodo feugiat nisl pulvinar pellentesque. Etiam auctor sodales
ligula, non varius nibh pulvinar semper. Suspendisse nec lectus non
ipsum convallis congue hendrerit vitae sapien. Donec at laoreet
eros. Vivamus non purus placerat, scelerisque diam eu, cursus
ante. Etiam aliquam tortor auctor efficitur mattis.

\section{Online Resources}

Nam id fermentum dui. Suspendisse sagittis tortor a nulla mollis, in
pulvinar ex pretium. Sed interdum orci quis metus euismod, et sagittis
enim maximus. Vestibulum gravida massa ut felis suscipit
congue. Quisque mattis elit a risus ultrices commodo venenatis eget
dui. Etiam sagittis eleifend elementum.

Nam interdum magna at lectus dignissim, ac dignissim lorem
rhoncus. Maecenas eu arcu ac neque placerat aliquam. Nunc pulvinar
massa et mattis lacinia.

\fi

\end{document}